\documentclass[11pt]{article}
\usepackage[,left=3cm,top=3cm,bottom=3cm,right=2.8cm,portrait]{geometry}
\usepackage{amsmath}
\usepackage{amssymb}
\usepackage{amsmath}
\usepackage{verbatim} 

\usepackage{graphicx}           
\usepackage{psfrag}               

\usepackage{kbordermatrix}      

\usepackage{array}
\usepackage{delarray}                 
\newcolumntype{L}{>{$}l<{$}}    
\newcolumntype{R}{>{$}r<{$}}
\newcolumntype{C}{>{$}c<{$}}

\newcommand{\beq}{\begin{equation*}}
\newcommand{\eeq}{\end{equation*}}
\newcommand{\beqn}{\begin{equation}}
\newcommand{\eeqn}{\end{equation}}
\newcommand{\bil}{\Big[\hspace{-0.225cm}\Big[\,}
\newcommand{\bir}{\,\Big]\hspace{-0.22cm}\Big]}

\newcommand{\p}{\partial}

\newcommand{\bs}[1]{\boldsymbol{#1}}
\newcommand{\D}{\displaystyle}

\newcommand{\bmb}{\begin{bmatrix}}
\newcommand{\bme}{\end{bmatrix}}

\begin{document}
\title{Reduced Effective Model for Condensation in Slender Tubes with Rotational Symmetry,
Obtained by Generalized Dimensional Analysis}
\author{Andrea Dziubek}
\date{}
\maketitle

\begin{abstract}
Experimental results for condensation in compact heat exchangers show that the heat transfer due to condensation is significantly better compared to classical heat exchangers, especially when using R134a instead of water as the refrigerant. This suggests that surface tension plays a role. Using generalized dimensional analysis we derive reduced model equations and jump conditions for condensation in a vertical tube with cylindrical cross section. Based on this model we derive a single ordinary differential equation for the thickness of the condensate film as function of the tube axis. Our model agrees well with commonly used models from existing literature. It is based on the physical dimensions of the problem and has greater geometrical flexibility.
\end{abstract}

\section{Introduction}

Condensation is important in the refrigeration, automotive and process industries, \cite{Carey}. Higher energy efficiency requirements and the move to more environmentally friendly refrigerants increased the need for highly efficient heat transfer for in-tube condensation (and evaporation) processes, \cite{Satish}. Improved heat transfer technologies are currently used to save not only energy but rather to save space. Over the last decades experimental studies show that the heat transfer is better in compact heat exchangers than in classical tube condensers, which made compact heat exchangers popular, \cite{Collier}.

The hydrodynamic flow channels in such condensers have diameters in the millimeter range and are often inclined to the vertical.
The fundamental mechanisms of heat and mass transfer as well as of two phase flow in these small channels are not well understood.
In this paper we study mainly condensation in a vertical tube and make some comments on inclined tubes.\\

Literature on condensation in tubes (or channels) with small diameters is on vertical,
inclined, or horizontal tubes (or channels) and the effect of surface tension is either taken
into account or not.

Condensation in vertical tubes is investigated by the following authors, where only the first
two authors considered surface tension. According to \cite{WangDuVertical} small surface
waves enhance the heat transfer mainly due to film thinning effect. \cite{Zhao} investigated
condensation in vertical triangular channels with a diameter between 0.2~mm and 0.3~mm.
\cite{Pan} showed that the effect of interfacial shear stress on the heat transfer depend on
the vapor velocity and on the mass transfer. \cite{Panday} showed that turbulent flow
enhances the heat transfer (in a tube with a diameter of 24~mm).

Some studies on inclined tubes are related to our study. \cite{Fieg} derived an analytical
solution for condensation in and on elliptical cylinders. He neglected variations of the film
thickness with the tube radius. \cite{Mosaad}~studied interfacial shear stress without
considering surface tension. \cite{WangDuInclined} compared heat transfer in horizontal
and inclined tubes with constant diameter by energy considerations.

\cite{Siow} studied condensation in horizontal parallel plate channels without considering
surface tension. \cite{WangHonda} investigated horizontal micro-fin tubes by dividing the flow in
two flow regimes.

Heat exchangers often operate at moderate Reynolds numbers ($Re\leq 100$), with a film
thickness in the millimeter range and small surface waves with wave length in the
centimeter range. It is well known from linear stability analysis of thin films, that surface
tension has a stabilizing effect on the film, \cite{Benjamin}, \cite{Yih}, and that
condensation (opposed to evaporation) tends to stabilize the film, \cite{Bankhoff},
\cite{Marshall}, \cite{Unsal}, \cite{Spindler}. Nonlinear stability analysis for flows along
plates has shown that stability of the film depends on the frequency of the initial
perturbation, see \cite{Burelbach}, \cite{Joo} (evaporation), and \cite{Hwang} (condensation).\\

In most of these studies simplified model equations were used. Interestingly, sometimes different equations for the interface between the two phases (jump conditions) are used. Often the model equations are simplified intuitively or the method of simplification remains unclear. In this paper we derive effective model equations for condensation in vertical tubes with cylindrical cross sections where surface tension is taken into account. We simplify the model equations using generalized dimensional analysis (GDA), an extension of dimensional analysis that allows to assign different length scales for each spatial variables. This method has mathematical rigor and is very algorithmic, but not well documented in literature, so we review it in this study.\\

The rest of the paper is organized as follows: In section \ref{mathmodel} the equations for the condensate and vapor phase and for the interface between the two phases are given. In section \ref{gda} we review generalized dimensional analysis (GDA). In section \ref{GDAapplied} and \ref{GDAjump} we obtain reduced model equations for the condensate flow, the vapor phase and the jump conditions between the two phases, based on the ratio of the film thickness and the tube length, $\varepsilon=H/L$, and we obtain the nondimensional numbers of the problem.
In section \ref{odesection} we evaluate the dimensionless numbers for water and R134a and obtain a single ordinary differential equation for the film thickness and in section \ref{solutions} we compute the heat transfer and compare our results to commonly used models from the literature.


\section{Mathematical model \label{mathmodel}}

 We consider a vertical tube with cylindrical cross section so that gravity acts in the direction of the tube
 axis and we assume laminar condensate flow. As the velocity of the condensate flow increases the
interface between condensate and vapor becomes wavy with at first two-dimensional surface
waves, for small surface waves the condensate flow is still laminar, \cite{Alekseenko},
\cite{Yoshimura}.  Based on this we can assume rotational symmetry so that the variables
do not depend on the rotation angle.\\

\begin{figure}[h!]
\psfrag{n}{$\bs{n}$}\psfrag{t}{$\bs{t}$}\psfrag{z}{$z$}\psfrag{r}{$r$}\psfrag{R}{$R$}\psfrag{D}{$H$}
\centering \leavevmode
\includegraphics[width=0.3\textwidth]{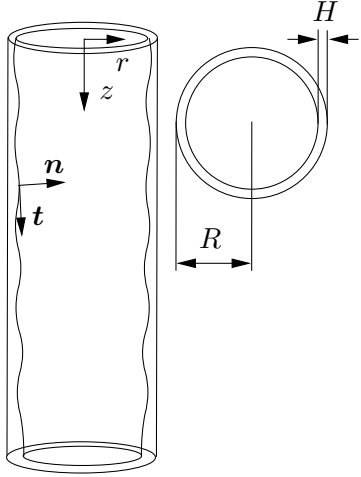}
\caption{Tube with rotational symmetry\label{tube}}
\end{figure}

In cylindrical coordinates the \textbf{domain of the condensate flow} is given by \beq
\Omega_{cond}=\{ (r(t) \cos \vartheta, r(t) \sin \vartheta, z) \in \mathbb R^3 \: |\: r(t)\in
[h(z,t),R), \; \vartheta \in [0,2\pi), \; z\in (0,L) \} \;,\eeq and the \textbf{domain of the
vapor} is given by \beq \Omega_{vapor}=\{ (r(t) \cos \vartheta, r(t) \sin \vartheta, z)\in
\mathbb R^3 \: |\: r(t)\in (0,h(z,t)), \; \vartheta \in [0,2\pi), \; z\in (0,L) \}\;, \eeq where
$h(z,t)$ is the thickness of the vapor phase, $R$ is the tube diameter and $L$ is the tube
length. We denote the thickness of the condensate film by
$H(z,t)=R-h(z,t)$.\\

Assuming that the condensate and the vapor can be  modeled as an incompressible
Newtonian fluid the \textbf{mass, momentum, and energy
balance equations} for the condensate flow in cylindrical coordinates are given by %
\begin{align}
 \label{cyl_konti}
\frac{1}{r}\,\frac{\p }{\p r}\left( r\,v_r \right) + \frac{\p v_z}{\p
z}=0\;, \\[1ex]
\label{cyl_nvs_rad}
\rho\,\left( \frac{\p v_r}{\p t}  + v_r\,\frac{\p v_r}{\p r}  + v_z\frac{\p
v_r}{\p z} \right) = - \frac{\p p}{\p r}  + \mu\,\left(\frac{\p}{\p
r}\left[\frac{1}{r}\,\frac{\p}{\p r}(r\,v_r) \right]
 + \frac{\p^2 v_r}{\p z^2} \right) \,, 
\end{align}
\begin{align}
 \label{cyl_nvs_ax}
\rho\,\left( \frac{\p v_z}{\p t} + v_r\,\frac{\p v_z}{\p r}
 + v_z\,\frac{\p v_z}{\p z} \right)
= - \frac{\p p}{\p z}
 + \mu\,\left( \frac{1}{r}\frac{\p}{\p r}\left[r\frac{\p v_z}{\p r}\right]
    + \frac{\p^2 v_z}{\p z^2}\right) + \rho\,g \; ,\\[1ex]
\label{cyl_eb} \rho\,c\,\left( \frac{\p T}{\p t} + v_r\, \frac{\p T}{\p r} + v_z\,\frac{\p T}{\p
z} \right) = \lambda\,\left( \frac{1}{r}\,\frac{\p}{\p r}\left[r\,\frac{\p T}{\p r}\right] +
\frac{\p^2 T}{\p z^2} \right) \;,
\end{align}%
where $\bs{v}=\left[v_r(r,z,t), v_z(r,z,t)\right]^T$ is the condensate velocity, $p(r,z,t)$ is the
condensate pressure, $T(r,z,t)$ is the condensate temperature, and  $g$ is the gravity. We
assume constant density $\rho$, dynamic viscosity $\mu$, specific heat capacity $c$, and
thermal conductivity  $\lambda$ of the condensate.\\

The velocity of the vapor phase is assumed to be small, so that shear stress exerted by the
vapor on the condensate film can be neglected. The vapor temperature shall be constant, i.e.\
we assume zero heat flux in the vapor phase. The pressure in the vapor phase is determined
mainly by the hydrostatic pressure,
 \beqn \label{nvs_gas}
 \frac{\p p_v}{\p z} = \rho_v\,g \;,
\eeqn%
 \cite{Fiedler}. Under this conditions the problem reduce to a single-phase problem and only the equations for
the condensate phase have to be solved. In the following we refer to equations
\eqref{cyl_konti} -- \eqref{nvs_gas} as the bulk equations.\\

The \textbf{boundary conditions} for the momentum equations are: non-homogeneous
Dirichlet boundary condition at the tube inlet, \beqn \label{bc_vp_in} v_z(r,z=0,t) = U(r)\;,
\eeqn
zero velocity at the tube wall,\\
\begin{minipage}[o]{0.45\textwidth}
\beqn \label{bc_vp_wall} v_r(r=R,z,t) = 0\;, \eeqn
\end{minipage}
\begin{minipage}[o]{0.45\textwidth}
\beqn \label{bc_vp_wall.2}  v_z(r=R,z,t) = 0\;,\eeqn
\end{minipage}\\[1.5ex]
and the outflow condition at the tube outlet,\\
\begin{minipage}[o]{0.45\textwidth}
\beqn \label{bc_vp_out}  v_r(r,z=0,t) = 0\;,\eeqn
\end{minipage}
\begin{minipage}[o]{0.45\textwidth}
\beqn \label{bc_vp_out.2}    \left(\frac{\p v_z}{\p z}\right)(r,z=L,t) = 0\;.\eeqn
\end{minipage}\\[1.5ex]
The boundary conditions for the energy equation are: the temperature at the tube wall is
known, \beqn  \label{bc_T_wall}
 T(r=R,z,t)=T_w \;,
\eeqn
zero temperature gradient normal to the tube inlet and outlet,\\
\begin{minipage}[o]{0.45\textwidth}
\beqn  \label{bc_T_in} \left(\frac{\p T}{\p z}\right)(r,z=0,t) = 0\;,\eeqn
\end{minipage}
\begin{minipage}[o]{0.45\textwidth}
\beqn  \label{bc_T_out} \left(\frac{\p T}{\p z}\right)(r,z=L,t) = 0\;,\eeqn
\end{minipage}\\[1.5ex]
and we assume thermodynamic equilibrium, i.e.\ the temperature at the interface is the
saturation temperature of the vapor, \beqn \label{bc_T_pgf} T(r=h(z,t),z,t) = T_s\;. \eeqn
Pressure boundary conditions will be discussed in section~\ref{GDAapplied}.\\

The \textbf{mass, momentum, and energy jump conditions} for the interface between
condensate and vapor are with the assumption of no-slip at the interface, $\bil
\bs{v}\cdot{\bs{t}}\bir=0$, constant surface tension $\gamma$, and after neglecting kinetic
and viscous terms in the energy jump condition given as
\begin{align} \label{mass-jump}
\bil \rho\,\left[\bs{v}-\bs{u}\right]\cdot{\bs{n}} \bir &= 0 \;,\\
\bil \dot{m}\,\bs{v}\cdot{\bs{n}} \bir + \bil p \bir - \bil {\bs{n}}\cdot\bs{D}\cdot{\bs{n}} \bir &=  2\,H\,\gamma \;, \\
\bil {\bs{t}}\cdot\bs{D}\cdot{\bs{n}} \bir &= 0 \;, \\
\dot{m}\Delta h_v + \bil \bs{q}\cdot{\bs{n}} \bir &= 0 \;, \label{energy-jump}
\end{align}
where $\bs{u}$ is the velocity of the interface, $\dot{m}=
\rho\,\left[\bs{v}-\bs{u}\right]\cdot{\bs{n}} $ is the volume specific mass flux,
$\bs{D}=\mu\left[\bs{\nabla}\bs{v}+(\bs{\nabla}\bs{v})^T\right]$ is the deviatoric stress
tensor for an incompressible Newtonian fluid, $H=-\tfrac{1}{2}\bs{\nabla}_S\cdot\bs{n}$ is
the mean curvature ($\bs{\nabla}_S$ is the surface gradient), $\Delta h_v$ is the latent
heat of vaporization, $\bs{q}=\lambda\bs{\nabla}T$ is the heat flux vector, and $\bs{n},
\bs{t}$ are the normal and tangent vector respectively. For a more detailed discussion of
jump conditions see \cite{Dziubek}.

Parameterizing the interface implicitly by $F(r,z,t)=r-h(z,t)=0$ we obtain from
\eqref{mass-jump} -- \eqref{energy-jump}
\begin{align}
\label{cyl_konti_pgf}
\dot{m} = \frac{\rho}{ \sqrt{ 1+ \left(\frac{\p h}{\p z}\right)^2}}
           \left( -v_r+\frac{\p h}{\p t}+v_z\,\frac{\p h}{\p z} \right)\;,\\[1ex]%
\label{cyl_nvs_pgf_n}
\dot{m} \left( -v_r + v_z\,\frac{\p h}{\p z} \right) + p - p_v
-\frac{\mu}{ 1+\left(\frac{\p h}{\p z}\right)^2 }
   \left( \frac{\p v_r}{\p r} - 4\left[\frac{\p v_r}{\p z}
       + \frac{\p v_z}{\p r}\right]\frac{\p h}{\p z}
     + \frac{\p v_z}{\p z}\left[\frac{\p h}{\p z}\right]^2 \right) \nonumber\\
     = -\gamma\left( -\frac{\frac{\p^2 h}{\p z^2}}{ \left(1+\left(\frac{\p h}{\p
z}\right)^2\right)^{3/2} }
        + \frac{1}{h\sqrt{ 1+\left(\frac{\p h}{\p z}\right)^2} }
        \right)\;,\\[1ex]
\label{cyl_nvs_pgf_t} \frac{\mu}{ \sqrt{ 1+\left(\frac{\p h}{\p z}\right)^2  } }\left(
     - \frac{\p v_r}{\p r}\frac{\p h}{\p z}
     - 2\left[\frac{\p v_r}{\p z}+\frac{\p v_z}{\p r}\right]
                     \left[1-\left(\frac{\p h}{\p z}\right)^2\right]
     + \frac{\p v_z}{\p z}\,\frac{\p h}{\p z} \right)   = 0 \;,\\[1ex]%
\label{cyl_pgf_eb}  \dot{m}\,\Delta h_v=-\frac{\lambda}{ \sqrt{ 1+\left(\frac{\p h}{\p
z}\right)^2 } } \left( -\frac{\p T}{\p r} + \frac{\p T}{\p z}\,\frac{\p h}{\p z} \right)\;,
\end{align}

\section{Generalized dimensional analysis \label{gda} }

In classical dimensional analysis all spatial variables are scaled by the same length scale.
\textbf{Generalized dimensional analysis}  is an extension of dimensional analysis
which allows different length scales for each spatial variable, \cite{Neemann}.\\

Length, mass and time (LTM) are fundamental dimensions of the MKS unit system and the
c.g.s unit system. Meter, kilogram, second (MKS) are units and centimeter, gram, second
(c.g.s) are other units. Another set of fundamental dimensions are length, force and time
(LFT). An equation in which the units balance on both sides of the equal sign is called
\textbf{coherent}. An equation in which the dimensions are equal on both sides of the equal
sign is called \textbf{homogeneous}.\footnote{For example, the equation
$3\,\text{m}+3\,\text{cm}=3\,\text{m}+3\times 0.01\,\text{m}=3.03\,\text{m}$ is
homogeneous but not coherent.}\\

According to \textbf{Buckingham's $\Pi$-Theorem} every physically meaningful equation
$f(a_1,\dots,a_n)=0$ with $n$ variables,  where the $n$ variables are expressed in
terms of $r$ fundamental dimensions, can be rewritten as an equation of $n-r=k$
dimensionless variables constructed from the original variables,  $F(\pi_1,\pi_2,\dots,\pi_k)
= 0$. Physically meaningful equations have to be invariant
under a change of system of units and this is used in a dimensional analysis.\\

For a generalized dimensional analysis the variables in the differential equations and in the
boundary conditions are substituted by their product of value and dimension. Because every
equation must be of dimensional homogeneity this result in equations for the dimensions
only. The derivatives are not carried out over the dimensions.
For example, from the
continuity equation,
\beq \frac{1}{\hat{r}\tilde{r}}\frac{\p }{\p \hat{r}\tilde{r}}\left(
\hat{r}\tilde{r}\,\hat{v}_r\tilde{v}_r \right) +\frac{\p \hat{v}_z\tilde{v}_z}{\p
\hat{z}\tilde{z}} = 0 \;,
\eeq we get the  \textbf{normalized dimension equation}
$\tilde{r}^{-1}\,\tilde{z}\,\tilde{v}_r\,\tilde{v}_z^{-1} = 1$.\\ 

A system of physically meaningful ordinary or partial differential equations with $n$
variables, where the $n$ variables are expressed in terms of $r$ fundamental dimensions,
result in $l$ normalized dimension equations %
\begin{align} \label{koherent_generalized}
\tilde{a}_1^{b_{11}}\,\tilde{a}_2^{b_{12}}\,\dots\tilde{a}_n^{b_{1n}} &= 1\;,\notag\\
\tilde{a}_1^{b_{21}}\,\tilde{a}_2^{b_{22}}\,\dots\tilde{a}_n^{b_{2n}} &= 1\;,\notag\\
\vdots  \\
\tilde{a}_1^{b_{l1}}\,\tilde{a}_2^{b_{l2}}\,\dots\tilde{a}_n^{b_{ln}} &= 1\;. \notag
 \end{align}
The powers~$b_{ij}$ form the matrix
\beq
\bs{B} = \bmb b_{11} & b_{12} & \dots & b_{1n}\\
              b_{21} & b_{22} & \dots & b_{2n}\\
              \vdots & \vdots &       & \vdots\\
              b_{l1} & b_{l2} & \dots & b_{ln} \bme \;,
\eeq with $k=\text{rank}\,\bs{B}$. From the $l$ coherent equations $k$ are linear
independent, so that $k$ nondimensional variables can be constructed.

The dimension of  a variable $a_j$ is a function of power monomials of the dimensions of all
variables, so that we make the following ansatz:
\begin{align}  \label{power_product} %
\tilde{a}_1 &= \tilde{a}_1^{y_{11}}\,\tilde{a}_2^{y_{12}}\,\dots\tilde{a}_n^{y_{1n}}\;, \notag \\
\tilde{a}_2 &= \tilde{a}_1^{y_{21}}\,\tilde{a}_2^{y_{22}}\,\dots\tilde{a}_n^{y_{2n}}\;, \notag \\
&\vdots &\\
\tilde{a}_n &= \tilde{a}_1^{y_{n1}}\,\tilde{a}_2^{y_{n2}}\,\dots\tilde{a}_n^{y_{nn}}\;,
\notag
\end{align}
 where the coefficients $y_{ij}$ are unknown. Substituting the dimensions
$\tilde{a}_j$ in the dimension equations \eqref{koherent_generalized} by their power
monomials gives for the $i$th equation \beq \left(
\tilde{a}_1^{y_{11}}\,\tilde{a}_2^{y_{12}}\,\dots\tilde{a}_n^{y_{1n}}
  \right)^{b_{i1}}
\left( \tilde{a}_1^{y_{21}}\,\tilde{a}_2^{y_{22}}\,\dots\tilde{a}_n^{y_{2n}}
  \right)^{b_{i2}}\,\dots\,
\left( \tilde{a}_1^{y_{n1}}\,\tilde{a}_2^{y_{n2}}\,\dots\tilde{a}_n^{y_{nn}}
  \right)^{b_{in}} = 1\;,
\eeq and further  \beq \tilde{a}_1^{ b_{i1}\,y_{11} + b_{i2}\,y_{21} +\dots+ b_{in}\,y_{n1} }\,
\tilde{a}_2^{ b_{i1}\,y_{12} + b_{i2}\,y_{22} +\dots+ b_{in}\,y_{n2} }\,\dots\,\tilde{a}_n^{
b_{i1}\,y_{1n} + b_{i2}\,y_{2n} +\dots+ b_{in}\,y_{nn} } = 1 \;,\eeq
so that the exponents sum up to zero. Doing this for all $l$ unit equations result in the
following \textbf{system of linear equations} for the exponents of the dimensions: \beq
\bs{B}\,\bs{y}_j = \bs{0} \;. \eeq The Buckingham-$\Pi$ Theorem says that from
$k=\text{rank}\,\bs{B}$ linear independent equations, $k$ dimensionless variables can be
constructed. The number of fundamental dimensions to be prescribed are given by the
Buckingham $\Pi$-Theorem as the number of variables minus the number of
nondimensional variables, $n-k=r$.

We collect the $r$ column vectors corresponding to the fundamental dimensions in the
matrix $\bs{R}$ and the remaining $k$ column vectors in the matrix $\bs{K}$, so that
 \beq \bs{B} = [ \bs{K}\,\big|\,\bs{R} ] \;. \eeq
Note that the $r$ fundamental dimensions have to be chosen such that $\text{rank}\,\bs{K}
= \text{rank}\,\bs{B}$. Then the $k$
\textbf{nondimensional variables} are given by%
\beq\pi_j = \tilde{a}_1^{y_{1j}}\,\tilde{a}_2^{y_{2j}}\,\dots\,\tilde{a}_n^{y_{nj}}\;,
  \qquad\qquad j=1,\dots,k\;.
\eeq

If we assign different units (length scales) to the spatial dimensions the equations are not
any more coherent and the system $\bs{B}\bs{y}_i=\bs{0}$ becomes inconsistent. This is the
 interesting case. Then we have with GDA a rigorous method to determine the terms that
have to be dropped from the model equations in order to recover a consistent system and to
determine the characteristic dimensionless numbers of the problem. For multiscale problems
with several small parameters the system may have more than one solution.

\section{Reduced equations and dimensionless numbers of  condensate flow and vapor\label{GDAapplied}}

To obtain the \textbf{normalized dimension equations of the bulk equations} we start by
writing all independent and dependent variables, material constants, and constants that
appear in the boundary conditions as product of value (hatted variables) and dimension
(tilded variables),
\begin{align*}
r &= \hat{r}\,\tilde{r}\;, & v_r  &= \hat{v}_r\,\tilde{v}_r\;,  & \rho &= \hat{\rho}\,\tilde{\rho}\;, & R &= \hat{R}\,\tilde{R}\;,  &     U &= \hat{U}\,\tilde{U}\;, \\
z &= \hat{z}\,\tilde{z}\;, & v_z &= \hat{v}_z\,\tilde{v}_z\;, & \mu &= \hat{\mu}\,\tilde{\mu}\;,  & H &= \hat{H}\,\tilde{H}\;,  & T_w &= \hat{T}_w\,\tilde{T}_w\;,\\
t &= \hat{t}\,\tilde{t}\;, &     p &= \hat{p}\,\tilde{p}\;,         &        g &= \hat{g}\,\tilde{g}\;,              & L &= \hat{L}\,\tilde{L}\;,   & T_w &= \hat{T}_w\,\tilde{T}_w\;,\\%
&                                          & T    &= \hat{T}\,\tilde{T}\;,        &  c &= \hat{c}\,\tilde{c}\;, \\
&                                          & p_v &= \hat{p}_v\,\tilde{p}_v\;, & \lambda &= \hat{\lambda}\,\tilde{\lambda}\;.   
\end{align*}
Substituting the variables in the continuity equation~\eqref{cyl_konti} by the product of
value and dimension gives the normalized dimension equation \beqn \label{unit_konti}
\tilde{r}^{-1}\,\tilde{z}\,\tilde{v}_r\,\tilde{v}_z^{-1} = 1\,. \eeqn
From the momentum equations \eqref{cyl_nvs_rad} and \eqref{cyl_nvs_ax} we obtain\\
\begin{minipage}[o]{0.45\textwidth}
\begin{align}
 \tilde{r}\,\tilde{t}^{-1}\,\tilde{v}_r^{-1} & = 1 \;,
\label{1.nvs-t-konv} \\
\tilde{r}^{-1}\,\tilde{z}\,\tilde{v}_r\,\tilde{v}_z^{-1} & =  1 \;,
\label{1.nvs-konv-konv} \\
\tilde{r}\,\tilde{z}^{-1}\,\tilde{v}_r\,\tilde{v}_z\,\tilde{p}^{-1}\,\tilde{\rho}&= 1\;,
\label{1.nvs-konv-p}    \\
\tilde{r}\,\tilde{v}_r^{-1}\,\tilde{p}\,\tilde{\mu}^{-1} &= 1 \;,
\label{1.nvs-p-eta}\\
\tilde{r}^{-2}\,\tilde{z}^2 & =1\;,
\label{1.nvs-eta-eta2}\\
\nonumber
\end{align}
\end{minipage}\qquad\qquad
\begin{minipage}[u]{0.45\textwidth}
\begin{align}
\tilde{r}\,\tilde{t}^{-1}\,\tilde{v}_r^{-1} &= 1\;, \label{2.nvs-t-konv}\\
\tilde{r}^{-1}\,\tilde{z}\,\tilde{v}_r\,\tilde{v}_z^{-1} & = 1 \;, \label{2.nvs-konv-konv}\\
\tilde{v}_z^2\,\tilde{p}^{-1}\,\tilde{\rho} &= 1\;, \label{2.nvs-konv-p} \\
\tilde{r}^2\,\tilde{z}^{-1}\,\tilde{v}_z^{-1}\,\tilde{p}\,\tilde{\mu}^{-1} &= 1 \;, \label{2.nvs-p-eta}  \\
 \tilde{r}^{-2}\,\tilde{z}^2 & =  1 \;, \label{2.nvs-eta-eta2}\\
\tilde{z}^{-2}\,\tilde{v}_z\,\tilde{\rho}^{-1}\,\tilde{\mu}\,\tilde{g}^{-1} &= 1 \;.
\label{2.nvs-eta-g} 
\end{align}
\end{minipage}\\[1.5ex]
From the energy equation \eqref{cyl_eb} we obtain\\
\begin{minipage}[o]{0.45\textwidth}
\begin{align}
\tilde{r}\,\tilde{t}^{-1}\,\tilde{v}_r^{-1} & = 1\;,  \label{eb-t-konv}\\
\tilde{r}^{-1}\,\tilde{z}\,\tilde{v}_r\,\tilde{v}_z^{-1} & = 1\;, \label{eb-konv-konv}
\end{align}
\end{minipage}\qquad\qquad
\begin{minipage}[o]{0.45\textwidth}
\begin{align}
\tilde{r}^2\,\tilde{z}^{-1}\,\tilde{v}_z\,\tilde{\rho}\,\tilde{c}\,\tilde{\lambda}^{-1} &= 1\;, \label{eb-konv-lambda} \\
\tilde{r}^{-2}\,\tilde{z}^2 &=  1  \label{eb-lambda-lambda2} \;.
\end{align}
\end{minipage}\\[1.5ex]
From the momentum equation of the vapor phase \eqref{nvs_gas} we obtain \beq
\label{unit_nvs_gas} \tilde{z}^{-1}\,\tilde{p}_v\,\tilde{\rho}_v^{-1}\,\tilde{g}^{-1} = 1\;.
\eeq
\eqref{unit_konti}, \eqref{1.nvs-konv-konv}, \eqref{2.nvs-konv-konv},\eqref{eb-konv-konv}
are identical, \eqref{1.nvs-t-konv}, \eqref{2.nvs-t-konv}, \eqref{eb-t-konv} are identical, and
\eqref{1.nvs-eta-eta2}, \eqref{2.nvs-eta-eta2}, \eqref{eb-lambda-lambda2} are identical.
From the four pressure related dimension equations \eqref{1.nvs-konv-p},
\eqref{1.nvs-p-eta}, \eqref{2.nvs-konv-p}, \eqref{2.nvs-p-eta} two are linear independent, but
before we eliminate linear dependent equations we analyze the boundary
conditions and study its consequences.\\

Now we obtain the \textbf{dimension equations of the boundary conditions}. From the
non-homogeneous Dirichlet boundary condition in terms of values and dimensions,
\eqref{bc_vp_in}, \beq
\hat{v}_z\,\tilde{v}_z\left(\hat{r}\,\tilde{r},\hat{z}\,\tilde{z}=0,\hat{t}\,\tilde{t}\right) =
\hat{U}\,\tilde{U}(\hat{r}\,\tilde{r})\;, \eeq we obtain that the streamwise velocity has the
same dimension as the velocity at the inlet, %
\beqn \label{unit_bc_vU} \tilde{v}_z = \tilde{U} \;. \eeqn The dimension equation
$\tilde{z}=0$ is homogeneous and provides no information.

From the remaining velocity boundary conditions at the wall \eqref{bc_vp_wall},
\eqref{bc_vp_wall.2} and the outflow condition \eqref{bc_vp_out.2} we obtain \\
\begin{minipage}[o]{0.45\textwidth}
\beqn \label{unit_bc_rH} \tilde{r} = \tilde{R}\;,\eeqn
\end{minipage}
\begin{minipage}[o]{0.45\textwidth}
\beqn \label{unit_bc_zL} \tilde{z} = \tilde{L}\;.\eeqn
\end{minipage}\\[1.5ex]
From the energy boundary conditions at the wall \eqref{bc_T_wall} and at the interface
\eqref{bc_T_pgf} together with with $h(z,t)=R-H(z,t)$ we obtain,\\
\begin{minipage}[o]{0.25\textwidth}
\beqn \label{unit_bc_TTw} \tilde{T} = \tilde{T}_w\;, \eeqn
\end{minipage}
\begin{minipage}[o]{0.25\textwidth}
\beqn \label{unit_bc_hR}\tilde{h} = \tilde{R} \;,\eeqn
\end{minipage}
\begin{minipage}[o]{0.25\textwidth}
\beqn  \label{unit_bc_RD} \tilde{R} = \tilde{H}\;,\eeqn
\end{minipage}
\begin{minipage}[o]{0.25\textwidth}
\beqn \label{unit_bc_TTs} \tilde{T} = \tilde{T}_s\;.\eeqn
\end{minipage}\\[1.5ex]
The inlet and outlet boundary conditions of the energy equation provide no new
information.\\

Now, we \textbf{assign tube radius and tube length to the spatial dimensions},
$\tilde{R}=\tilde{H}=H$ and $\tilde{L}=L$. However, with this length scales (units) the linear
equation system becomes inconsistent. To recover a consistent system of dimension
equations we compare the order of the terms and drop higher order terms.

\begin{itemize}
\item[i)] The second derivatives in streamwise direction are of order~$\varepsilon^2=(H/L)^2$
smaller than the second derivatives in radial direction and should be dropped, so that
\eqref{1.nvs-eta-eta2}, \eqref{2.nvs-eta-eta2}, and \eqref{eb-lambda-lambda2} vanish. Note
that eliminating \eqref{2.nvs-eta-eta2} changes \eqref{2.nvs-eta-g} to \beqn
\label{2.nvs-eta-g-changed}
\tilde{r}^{-2}\,\tilde{v}_z\,\tilde{\rho}^{-1}\,\tilde{\mu}\,\tilde{g}^{-1} =1\; \eeqn
\item[ii)] The pressure related dimension equations \eqref{1.nvs-konv-p}, \eqref{1.nvs-p-eta} and
\eqref{2.nvs-konv-p}, \eqref{2.nvs-p-eta} contradict dimension equation \eqref{unit_konti}.
Setting the radial derivative of the pressure to zero solves the problem. Then the dimension
equations~\eqref{1.nvs-t-konv} -- \eqref{1.nvs-p-eta} vanish.
\item[iii)] Consequently the condensate pressure is a function of streamwise coordinate and time only $
p = p(z,t)$ and is determined by the hydrostatic pressure of the vapor $ \tfrac{\p p}{\p
z}=\tfrac{\p p_v}{\p z} =\rho_v\,g$. The radial velocity is one order smaller compared to
streamwise velocity.\\
\end{itemize}

\textbf{Remark 1.} We see that i) -- iii) are results of our generalized dimensional analysis.
However, when Prandl derived his famous boundary layer equations he used i) -- iii) as initial assumptions, \cite{Schlichting}.
(Prandtl had a great intuition.)\\

The boundary conditions also determine the dimensions of vapor thickness and film
thickness, $\tilde{h}=H$ and $\tilde{D}=H$.  For simplicity we further assign the dimension
of inlet velocity and saturation temperature, $\tilde{U}=U$ and $\tilde{T}_s=T_s$. Then the
dimension of temperature and wall temperature are $\tilde{T}_w=T_s$ and
$\tilde{T}=T_s$.\\

With the simplifications discussed above and after eliminating linear dependent equations
the dimension equations form a system of $7$ homogeneous linear equations for~$n=12$
unknown dimensions, $\bs{B}\,\bs{x}=\bs{0}$, with $k=\text{rank}\,\bs{B}=7$, so that
$\bs{B}$ has maximum rank. From this equation system $k=7$ nondimensional variables
can be constructed from $r=n-k=5$ fundamental dimensions, which have to be chosen such
that the matrix $\bs{R}$  in $\bs{B}=[\bs{K}\big|\bs{R}]$ fulfills the condition
$\text{rank}\,\bs{K} = \text{rank}\,\bs{B}$.

Some dimensions are already determined by the boundary conditions. In addition to the
spatial dimensions and the dimension of the inlet velocity we chose the dimensions of the
two material properties density and heat capacity as fundamental dimensions, so that \beq
\hspace{-0.5cm} \bs{B}=
\kbordermatrix{%
& \tilde{v_r} & \tilde{p} & \tilde{t} &
    \tilde{\mu} &  \tilde{g} & \tilde{\lambda} & \tilde{\rho}_v &\;\;
   \tilde{\rho} & \tilde{c} & \tilde{v}_z & \tilde{r} & \tilde{z} \cr
\eqref{unit_konti}           &  1&   &   &   &   &   &  &\omit\vrule\;\;   &  & -1& -1& 1 \cr
\eqref{2.nvs-t-konv}         & -1&   & -1&   &   &   &  &\omit\vrule\;\;   &  &   & 1 &    \cr
\eqref{2.nvs-konv-p}         &   & -1&   &   &   &   &  &\omit\vrule\;\; 1 &  & 2 &   &    \cr
\eqref{2.nvs-p-eta}          &   &  1&   & -1&   &   &  &\omit\vrule\;\;   &  & -1& 2 & -1 \cr
\eqref{2.nvs-eta-g-changed}  &   &   &   & 1 & -1&   &  &\omit\vrule\;\;-1 &  & 1 & -2&    \cr
\eqref{eb-konv-lambda}       &   &   &   &   &   & -1&  &\omit\vrule\;\; 1 & 1& 1 & 2 & -1 \cr
\eqref{unit_nvs_gas}         &   & 1 &   &   & -1&   &-1&\omit\vrule\;\;   &  &   &   & -1 }
\negthinspace. \eeq

Setting $\tilde{\rho}=\rho$, $\tilde{c}=c$ we obtain with $\tilde{r}=H$, $\tilde{l}=L$,
$\tilde{v}_z=U$ after applying Gau{\ss}ian elimination to~$\bs{B}=[\bs{K}\big|\bs{R}]$
from the row-reduced echelon form of $\bs{B}$ the following \textbf{nondimensional
variables:}
\begin{gather*}
\pi_1 = \frac{\tilde{v}_r\,L}{U\,H}\;,\qquad 
\pi_2 = \frac{\tilde{p}}{\rho\,U^2}\;,\qquad  
\pi_3 = \frac{\tilde{t}\,U}{L}\;,\\
\pi_4 = \frac{\tilde{\mu}\,L}{\rho\,U\,H^2}\;,\qquad
\pi_5 = \frac{\tilde{g}\,L}{\tilde{U}^2}\;,\qquad  
\pi_6 = \frac{\tilde{\lambda}\,L}{\rho\,c\,U\,H^2}\;,\qquad
\pi_7 = \frac{\tilde{\rho}_v}{\rho}\;. 
\end{gather*}
Defining Reynolds, Froude, Prandtl and Peclet numbers, \beq \text{Re}
=\frac{\rho\,U\,H}{\mu}\;,\qquad \text{Fr} =\frac{U^2}{g\,H}\;,\qquad \text{Pr}
=\frac{\mu\,c}{\lambda}\;,\qquad \text{Pe} =\frac{U\,H\,\rho\,c}{\lambda}\;, \eeq we
get with $\tilde{\mu}=\mu$, $\tilde{g}=g$ and $\tilde{\lambda}=\lambda$, \beq \pi_4
=\frac{1}{\varepsilon\,\text{Re}}\;,\qquad \pi_5 =
\frac{1}{\varepsilon\,\text{Fr}}\;,\qquad \pi_6 =
\frac{1}{\varepsilon\,\text{Re}\,\text{Pr}}= \frac{1}{\varepsilon\,\text{Pe}}\;. \eeq Using
this we obtain from~\eqref{cyl_konti} --\eqref{nvs_gas} the \textbf{reduced mass,
momentum, and energy equations} in dimensionless form as the following system of partial
differential equations:
\begin{align}
 \label{simple_konti} \frac{1}{\hat{r}}\,\frac{\p }{\p \hat{r}}\left(
\hat{r}\,\hat{v}_r\right)
        + \frac{\p \hat{v}_z}{\p \hat{z}} &= 0\;,\\[1ex]
\label{simple_nvs_1} \left[ \frac{\p \hat{v}_z}{\p \hat{t}} + \hat{v}_r\,\frac{\p
\hat{v}_z}{\p\hat{r}} + \hat{v}_z\,\frac{\p \hat{v}_z}{\p \hat{z}} \right] &=
-\frac{\p\hat{p}}{\p\hat{z}} +
\frac{1}{\varepsilon\,\text{Re}}\,\frac{1}{\hat{r}}\frac{\p}{\p\hat{r}}\left(\hat{r}\frac{\p
\hat{v}_z}{\p \hat{r}}\right) + \frac{1}{\varepsilon\,\text{Fr}}\;,\\[1ex]
\label{simple_nvs_2}
\frac{\p\hat{p}}{\p\hat{r}} &= 0 \;,\\[1ex]
\label{simple_eb} \left[ \frac{\p\hat{T}}{\p\hat{t}} + \hat{v}_r\,\frac{\p\hat{T}}{\p\hat{r}}
    + \hat{v}_z\,\frac{\p\hat{T}}{\p\hat{z}} \right] &=
    \frac{1}{\varepsilon\,\text{Pe}}\,\left( \frac{1}{\hat{r}}\,\frac{\p}{\p\hat{r}}
    \left[\hat{r}\,\frac{\p\hat{T}}{\p\hat{r}}\right]  \right) \;,
\end{align}
and the momentum equation of the vapor flow, \beqn \label{simple_nvs_gas}
\frac{\p\hat{p}_v}{\p\hat{z}} = \rho_v\,g\;. \eeqn These equations are the zero and first
order equations for the bulk flow.\\

\textbf{Remark 2.} Nusselt's theory of condensation along a flat plate is based on zero order
equations, \cite{Nusselt}. Our model is an extension of Nusselt's theory for condensation in a
tube of rotational symmetry.

\section{Reduced jump conditions and dimensionless numbers of the interface equations\label{GDAjump}}

Now we use GDA to derive reduced jump conditions. We have three additional variables,
\begin{align*}
 h &= \hat{h}\,\tilde{h}\;,  & \gamma &=\hat{\gamma}\,\tilde{\gamma}\;, & \Delta h_v &= \hat{\Delta h}\,\tilde{\Delta h}\;.
\end{align*}
First, we obtain the \textbf{dimension equations of the jump conditions}.  All jump conditions
\eqref{cyl_konti_pgf} -- \eqref{cyl_pgf_eb} contain the term $\sqrt{ 1+ \left(\tfrac{\p h}{\p
z}\right)^2}$. With $1\gg\varepsilon^2\left(\tfrac{\p\hat{h}}{\p\hat{z}}\right)^2$ this
term reduces to one, so that the square root in the jump conditions vanishes. Also the term
$\left( 1 -\left(\frac{\p h}{\p z}\right)^2 \right)$ in the tangential momentum jump condition vanishes.\\

\textbf{Remark 3.} This reflects the assumption of long wavelength approximation.\\

Substituting the variables in the jump conditions by the product of value and dimension
and using the nondimensional numbers $\pi_1$ to $\pi_4$ and $\pi_6$ we obtain the
following jump conditions:
\begin{align}
\label{mdot_reduced} \dot{m} = \rho\varepsilon U\,\left( -\hat{v}_r
+\frac{\p\hat{h}}{\p\hat{t}} + \hat{v}_z\frac{\p\hat{h}}{\p\hat{z}} \right) = \rho
\varepsilon U\widehat{\dot{m}}\;, 
\end{align}
\begin{align}
\rho \varepsilon^2 U^2\widehat{\dot{m}}\left(-\hat{v}_r +
    \hat{v}_z\frac{\p\hat{h}}{\p\hat{z}} \right)
    + \rho U^2\left(\hat{p} - \hat{p}_v\right) \notag\\
    - \rho \varepsilon^2 U^2\,\hat{\mu}\left(\frac{\p\hat{v}_r}{\p\hat{r}}
    - 4\left(\varepsilon^2\,\frac{\p\hat{v}_r}{\p\hat{z}}
    +\frac{\p\hat{v}_z}{\p\hat{r}} \right)\frac{\p\hat{h}}{\p\tilde{z}}
    +\varepsilon^2\,\frac{\p\hat{v}_z}{\p\hat{z}}\left(\frac{\p\hat{h}}{\p\hat{z}}\right)^2\right)\label{nvs_n_reduced}\\
= -\frac{\tilde{\gamma}}{H}\hat{\gamma}\left(
-\varepsilon^2\,\frac{\p^2\hat{h}}{\p\hat{z}^2} + \frac{1}{\hat{h}}\right)\;,\notag\\[1ex]
\label{nvs_t_reduced}
\rho \varepsilon U^2\hat{\mu} \left( - \varepsilon^2\,\frac{\p
\hat{v}_r}{\p \hat{r}}\frac{\p \hat{h}}{\p \hat{z}} - 2\, \left(
\varepsilon^2\,\frac{\p\hat{v}_r}{\p\hat{z}} +\frac{\p\hat{v}_z}{\p\hat{r}} \right)
   + \varepsilon^2\,\frac{\p\hat{v}_z}{\p\hat{z}}\frac{\p\hat{h}}{\p\hat{z}}\right) = 0\;,\\[1ex]
\label{eb_reduced} \rho \varepsilon U\widetilde{\Delta
h}_v\,\widehat{\dot{m}}\widehat{\Delta h}_v = -\rho \varepsilon U\,c T_s\hat{\lambda}
\left(  -\frac{\p\hat{T}}{\p\hat{r}}
+\varepsilon^2\,\frac{\p\hat{T}}{\p\hat{z}}\frac{\p\hat{h}}{\p\hat{z}}\right)\;.
\end{align} %

The {mass jump condition}  \eqref{mdot_reduced} provides no additional dimension equation.

The first three terms in the {normal momentum jump condition} \eqref{nvs_n_reduced} have the
dimensions $\rho \varepsilon^2 U^2$, $\rho U^2$, and $\rho \varepsilon^2 U^2$.
Momentum transport due to condensation (or evaporation) and viscous normal stress exerted
on the interface are of order $\varepsilon^2$ smaller compared to the pressure terms so that
from the left hand side only the pressure term remains. From the two curvature terms the
term related to surface waves is of order two smaller than the term related to the tube
diameter, so that the first curvature term vanishes.

The {tangential momentum jump condition} \eqref{nvs_t_reduced} reduces to the condition of no
shear stress at the interface. This yields only a homogeneous dimension equation.

The second term on the right hand side of the {energy jump condition} \eqref{eb_reduced} is of
order two smaller than the first term and vanishes.\\

We obtain two normalized dimension equations, one from the normal momentum
jump condition and one from the energy  jump condition,\\
\begin{minipage}[o]{0.45\textwidth}
\beqn  \label{jump_gamma} \tilde{h}\,\tilde{p}_v\,\tilde{\gamma}^{-1} = 1\;,\eeqn
\end{minipage}
\begin{minipage}[o]{0.45\textwidth}
\beqn  \label{jump_latentheat}
\tilde{h}\,\tilde{r}\,\tilde{z}^{-1}\,\tilde{v}_z\,\tilde{T}^{-1}\,\tilde{\rho}\,\tilde{\lambda}^{-1}\,\widetilde{\Delta
h}_v=1\;.\eeqn
\end{minipage}\\[1.5ex]
This gives two additional \textbf{dimensionless numbers from the jump conditions}, \beq
\pi_{8} = \frac{\tilde{\gamma}}{\rho\,U^2\,H}\;, \qquad\qquad \pi_{9} =
\frac{\widetilde{\Delta h}_v}{c\,T_s}\;. \eeq %
Defining Weber number and Stefan number as \beq
\text{We}=\frac{\rho\,U^2\,H}{\gamma}\;,\qquad\qquad \text{St}=\frac{c\,T_s}{\Delta
h}_v\;, \eeq we get \beq\pi_8 = \frac{1}{\text{We}}\;,
\qquad\qquad\pi_9=\frac{1}{\text{St}}\,.\eeq
\medskip
Using this we obtain from~\eqref{mdot_reduced} -- \eqref{eb_reduced} the \textbf{reduced mass,
normal and tangential momentum, and energy jump conditions}  in dimensionless form as
the following system of partial differential equations:
\begin{align}
\label{simple_pgf_mass} \dot{m} &= \rho\,\varepsilon\,U\,\left(
      -\hat{v}_r + \frac{\p \hat{h}}{\p \hat{t}}
      + \hat{v}_z\,\frac{\p \hat{h}}{\p \hat{z}} \right)\;,\\[1ex]
\label{simple_pgf_momentum} \hat{p}-\hat{p}_v &= -\frac{1}{\text{We}}\frac{1}{\hat{h}}\;,\\[1ex]
\label{simple_pgf_shearstress} \frac{\p \hat{v}_z}{\p \hat{r}} &= 0\;,\\[1ex]
\label{simple_pgf_heat} \frac{1}{\text{St}}\,\left( -\hat{v}_r + \frac{\p \hat{h}}{\p \hat{t}} +
\hat{v}_z\,\frac{\p \hat{h}}{\p \hat{z}} \right) &=
\frac{1}{\varepsilon\text{Pe}}\frac{\p \hat{T}}{\p \hat{r}} \;.%
\end{align}
The normal and tangential momentum jump conditions are zero order equations.
\eqref{simple_pgf_shearstress} is the condition of no shear stress at the interface.
Condensation is a first order process, i.e.\ mass flux terms and phase change terms in the
energy jump condition have order epsilon.\footnote{For condensation along a flat plate
curvature due to surface waves is the only curvature term and for higher Reynolds numbers
surface waves may become wavy and cause instabilities.}\\

\textbf{Remark 4.} \eqref{simple_pgf_momentum} is the Young-Laplace equation.

\section{One ordinary differential equation model\label{odesection}}

First, we compare the dimensionless numbers $\pi_4, \pi_5, \pi_6$ and $\pi_8, \pi_9$ for
water and R134a, a refrigerant widely used in automobile air conditioning, to motivate
further simplifications of the model equations. The dimensionless numbers are the
coefficients of the terms in the reduced balance equations~\eqref{simple_konti} --
\eqref{simple_nvs_gas} and in the reduced jump conditions~\eqref{simple_pgf_mass} --
\eqref{simple_pgf_heat}. See appendix \ref{material} for material values and constants. The
quotient of the two length scales is~$\varepsilon=0.2\,10^{-3}$.\\

\begin{center}
\begin{tabular}{|L|C|C|}
\hline
\textbf{dimensionless number} & \textbf{Water} & \textbf{R134a} \\
\hline
\qquad\pi_4=1/(\varepsilon\text{Re}) & 1282 & 58 \\
\qquad\pi_5=1/(\varepsilon\text{Fr}) & 8772 & 243 \\
\qquad\pi_6=1/(\varepsilon\text{Pe}) & 325 & 17 \\
\qquad\pi_8=1/\text{We} & 1250 & 3.37 \\
\qquad\pi_9=1/\text{St} & 0.002 & 0.42 \\
\hline
\end{tabular}\\[1ex]
\medskip
\end{center}

Clearly gravity forces and viscous forces are dominant compared to inertial and pressure
forces for both fluids. Conductive heat transport dominates convective heat transport --
considerably more for water than for R134a. The effect of surface tension is much greater for water
than for R134a, where the pressure is almost determined by the hydrostatic pressure. At the
interface heat conduction dominates condensation, again the difference is greater for water
than for R134a. Apparently the flow is the dominating process and the position of the moving
surface is mainly determined by the solution of the free surface problem.\\

Based on these considerations, we neglect transient and convective terms in the bulk equations and the
radial velocity in the mass-energy jump condition for the rest of the paper. Then the bulk
equations reduce to a system of two partial differential equations,
\begin{align}
\label{dgl_ST_nvs}
\frac{1}{\varepsilon\,\text{Re}}\,\frac{1}{\hat{r}}\frac{\p}{\p \hat{r}}
  \left(\hat{r}\frac{\p \hat{v}_z}{\p \hat{r}}\right)
&= \frac{\p \hat{p}}{\p \hat{z}} -\frac{1}{\varepsilon\,\text{Fr}}\;, \\
\label{dgl_ST_2}
\frac{\p}{\p\hat{r}}\left(\hat{r}\,\frac{\p\hat{T}}{\p\hat{r}}\right) &= 0\;.
\end{align}
together  with  $\hat{p}=\hat{p}(\hat{z})$, i.e.\ the pressure is the hydrostatic pressure.

Boundary conditions for velocity and temperature are prescribed at the wall and at the
interface,
\begin{align}
\label{bc_velocity}
\hat{v}_z(\hat{r}=\hat{R},\hat{z}) &= 0\;,&
\left(\frac{\p\hat{v}_z}{\p\hat{r}}\right)(\hat{r}=\hat{h},\hat{z}) &= 0\;,\\
\label{bc_temperature}
\hat{T}(\hat{r}=\hat{R},\hat{z}) &= \hat{T}_w\;,& \hat{T}(\hat{r}=\hat{h},\hat{z}) &=
\hat{T}_s\;.
\end{align}
 Setting $\hat{p}_v=0$ we have from the jump conditions at $\hat{r}=\hat{h}(\hat{z})$,%
\begin{align}
\label{dgl_ST_jump_nvs_n}
\hat{p}  &= -\frac{1}{We}\frac{1}{\hat{h}}\;,\\
\label{dgl_ST_3}
\hat{v}_z\,\frac{\p \hat{h}}{\p \hat{z}} &=
{\frac{\text{St}}{\varepsilon\,\text{Pe}}}\,\frac{\p \hat{T}}{\p \hat{r}}\;.
    \end{align}
The initial condition for the film thickness $\hat{h}$ is that the film thickness is prescribed
at the inlet, \beqn \label{bc_h0} \hat{h}(\hat{r},\hat{z}=0) = \hat{h}_0\;. \eeqn

Differentiating the normal momentum jump condition~\eqref{dgl_ST_jump_nvs_n} with
respect to~$\hat{z}$ and substituting the pressure gradient in~\eqref{dgl_ST_nvs} gives
\begin{align}
\label{dgl_ST_1}
\frac{1}{\hat{r}}\frac{\p}{\p \hat{r}}
    \left(\hat{r}\frac{\p \hat{v}_z}{\p \hat{r}}\right)
&= \frac{\varepsilon\,\text{Re}}{\text{We}}\,\frac{1}{\hat{h}^2}\,\frac{\p\hat{h}}{\p\hat{z}}
   -\frac{\text{Re}}{\text{Fr}}\;.
\end{align}
By integrating~\eqref{dgl_ST_1} and~\eqref{dgl_ST_2} and evaluating the boundary
conditions we obtain the dimensionless velocity as
\begin{align}
\label{vz_ST_ath}
\hat{v}_z &= \left(
-\underline{\frac{\varepsilon\text{Re}}{2\,\text{We}}\,\frac{1}{\hat{h}^2}\,\frac{\p\hat{h}}{\p\hat{z}}}
           + \frac{\text{Re}}{2\,\text{Fr}} \right)
    \left( \frac{\hat{R}^2-\hat{r}^2}{2}
           + \hat{h}^2\,\ln\frac{\hat{r}}{\hat{R}} \right)\;,
\end{align}
and the dimensionless temperature gradient as
\begin{align}
\label{T_sT_ath} \frac{\p \hat{T}}{\p \hat{r}} &=
\frac{\hat{T}_s-\hat{T}_w}{\ln(\hat{h}/\hat{R})}\,\frac{1}{\hat{r}}\;,
\end{align}
where the underlined term in \eqref{vz_ST_ath} represents the effect of surface tension and pressure in
\eqref{dgl_ST_nvs} and \eqref{dgl_ST_jump_nvs_n}.
Evaluating~\eqref{vz_ST_ath}
and~\eqref{T_sT_ath} at~$\hat{r}=\hat{h}(\hat{z})$ and substituting the results into the
mass-energy jump condition~\eqref{dgl_ST_3} gives
\begin{align} \label{ode_blaeh}
\left(
-\underline{\frac{\varepsilon\,\text{Re}}{2\,\text{We}}\,\frac{1}{\hat{h}^2}\,\frac{\p\hat{h}}{\p\hat{z}}}
  +\frac{\text{Re}}{2\,\text{Fr}} \right)
\left( \frac{\hat{R}^2-\hat{h}^2}{2}
  + \hat{h}^2\,\ln\frac{\hat{h}}{\hat{R}} \right)\,\frac{\p \hat{h}}{\p \hat{z}}
= \frac{\text{St}}{\varepsilon\,\text{Pe}}\,
  \frac{\hat{T}_s-\hat{T}_w}{\ln(\hat{h}/\hat{R})\,h}\;,
\end{align}
or equivalently
\beqn
\label{ode_ST}
\left(\frac{\p \hat{h}}{\p \hat{z}}\right)^2
  -\frac{\text{We}}{\varepsilon\text{Fr}}\,\hat{h}^2\,\frac{\p \hat{h}}{\p \hat{z}}
=-\frac{2\,\text{We}\,\text{St}\,(\hat{T}_s-\hat{T}_w)}{\varepsilon^2\,\text{Re}\,\text{Pe}}\,
   \frac{\hat{h}}{ \left( \frac{\hat{R}^2-\hat{h}^2}{2} +
                    \hat{h}^2\,\ln\frac{\hat{h}}{\hat{R}} \right)\ln(\hat{h}/\hat{R})}\;.
\eeqn
Note that with $\text{St}=\tfrac{c_p T_s}{\Delta h_v}$ and $\hat{T}_s-\hat{T}_w=1-\tfrac{T_s-\Delta T}{T_s}=\tfrac{\Delta T}{T_s}$ we have $\text{St}\,(\hat{T}_s-\hat{T}_w)=\tfrac{c_P \Delta T}{\Delta h_v}$. We call this new dimensionless number $ST$.\\

This equation is a fully nonlinear first order differential equation for the film
thickness. We write more conveniently,
\beq
\left(\frac{\p\hat{h}}{\p\hat{z}}\right)^2
  - a\,\hat{h}^2\,\frac{\p \hat{h}}{\p \hat{z}}
= -b\,f(\hat{h})\;,
\eeq
with~$a=\frac{\text{We}}{\varepsilon\,\text{Fr}}$
and~$b=\frac{2\,\text{We}\,\text{ST}}{\varepsilon^2\,\text{Re}\,\text{Pe}}$.
The left hand side can be transformed in an quadratic expression
\beq
\left( \frac{\p
\hat{h}}{\p \hat{z}} - a\,\frac{\hat{h}^2}{2} \right)^2
  - a^2\,\frac{\hat{h}^4}{4} = b\,f(\hat{h})\;,
\eeq
so that
\beq
\frac{\p \hat{h}}{\p \hat{z}} = a\,\frac{\hat{h}^2}{2} \pm
\sqrt{a^2\,\frac{\hat{h}^4}{4}-b\,f(\hat{h})}\;.
\eeq
Substituting back the parameters~$a$
and~$b$ we obtain the following quasilinear ordinary differential equation:
\beqn
\label{odeST}
\hspace{-0.3cm}
\frac{\p \hat{h}}{\p \hat{z}} =
\frac{\text{We}}{\varepsilon\,\text{Fr}}\,\frac{\hat{h}^2}{2}\stackrel{+}{-}\sqrt{
    \left(\frac{\text{We}}{\varepsilon\,\text{Fr}}\right)^2\,\frac{\hat{h}^4}{4}
    - \frac{2\,\text{We}\,\text{ST}}{\varepsilon^2\,\text{Re}\,\text{Pe}}\,
    \frac{\hat{h}}{ \left( \frac{\hat{R}^2-\hat{h}^2}{2}
    + \hat{h}^2\,\ln\frac{\hat{h}}{\hat{R}} \right) \ln\frac{\hat{h}}{\hat{R}} }}\;,
\eeqn
where the physical relevant solution describing condensation in a vertical tube with
small diameter is the solution with the negative signed root.\\

Neglecting the effect of surface tension and pressure in \eqref{ode_blaeh} gives a simpler
ordinary differential equation, \beqn \label{ode} \frac{\p \hat{h}}{\p \hat{z}} =
\frac{2\,\text{Fr}\,\text{ST}}{\varepsilon\,\text{Re}\,\text{Pe}} \,
  \frac{1}{ \left( \frac{\hat{R}^2-\hat{h}^2}{2}
            + \hat{h}^2\,\ln\frac{\hat{h}}{\hat{R}}\right)
           \ln\frac{\hat{h}}{\hat{R}} \, \hat{h} }\;.
\eeqn

\section{Numerical solutions and comparison with other models\label{solutions}}

Nusselt's equation for the dimensionless film thickness $\hat{H}$ of a condensate flow along a flat plate is a function of the dimensionless streamwise coordinate $\hat{z}$ to the power of one fourth, \cite{Bejan},
\beqn \label{odeNu} \hat{H}(\hat{z}) = \sqrt{2}\left( \frac{2\,\text{Fr}\,\text{ST}}{\text{Re}\,\text{Pe}}\,\hat{z} \right)^{\frac{1}{4}}\;.
\eeqn

\paragraph{Water}
We assume a tube diameter of 7~mm and take water as condensate, with an initial film thickness of $h_0=0.1$ \,mm. All computations are dimensionless.
Further details, such as wall temperature and the dimensionless numbers of water, can be found in the  appendix~\ref{material}. Although the model equations are nonlinear differential equations a standard Runge-Kutta(4) method is sufficient. We used \textit{Mathematica} for the computations.
\begin{figure}[h!]
\psfrag{Water}{Water}\psfrag{R-h}{$\hat{H}(\hat{z})$}\psfrag{z}{$\hat{z}$} \psfrag{bei}{\eqref{ode},\eqref{odeST}} \psfrag{nu}{Nu}
\centering
\leavevmode
\includegraphics[width=0.55\textwidth]{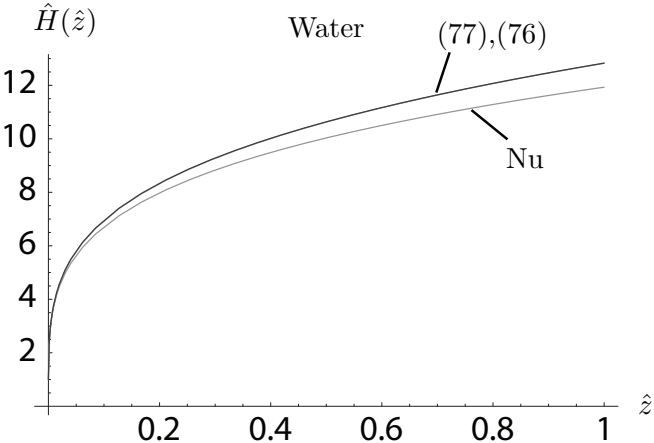}
\caption{ Water: Nu{\ss}elt \eqref{odeNu}, without surface tension \eqref{ode} with surface tension \eqref{odeST}\label{waterST}}
\end{figure}
The solutions for the model equations without surface tension and with surface tension are not distinguishable in figure~\ref{waterST}. The film thickness predicted by~\eqref{ode} and~\eqref{odeST} is slightly above the film thickness predicted by Nu{\ss}elt. The reason is that because of the circular tube the condensing mass result in a thicker condensate film.

\begin{figure}[h!]
\psfrag{Wasse}{Water} \psfrag{r}{}\psfrag{R-}{$\hat{H}$}\psfrag{h}{$(\hat{z})$} \psfrag{z}{$\hat{z}$} \psfrag{nu}{Nu} \psfrag{d8}{\hspace{-.1cm}$d\!=\!8$\,mm}\psfrag{d60}{\hspace{-.3cm}$d\!=\!60$\,mm}
\centering \leavevmode
\includegraphics[width=0.55\textwidth]{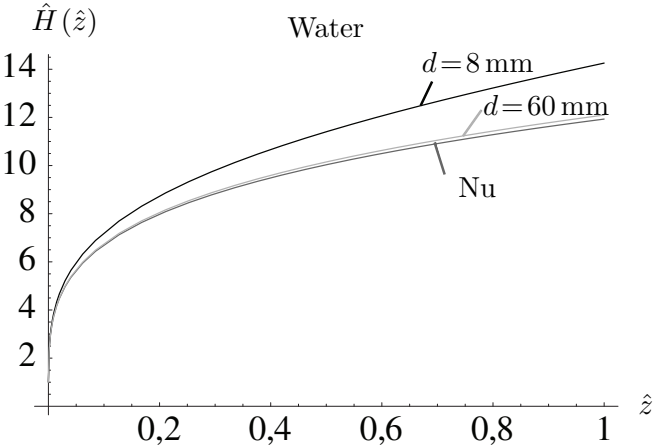}
\caption{Water: \eqref{odeNu}, \eqref{ode}, \eqref{odeST} for $d=60$~mm and $d=8$~mm
  \label{radii}}
\end{figure}

As a side effect we show in figure~\ref{radii} that for a diameter of about 60~mm the effect of surface tension due to the tube diameter becomes insignificant.\footnote{Note that the characteristic velocity and length are based on a diameter of 7 mm.}

\paragraph{R134a}

\begin{figure}[h!]
\psfrag{R134a}{R134a} \psfrag{R-h}{$\hat{H}(\hat{z})$} \psfrag{z}{$\hat{z}$}
\psfrag{bei}{\eqref{ode},\eqref{odeST}} \psfrag{nu}{Nu{\ss}elt} \centering
\leavevmode
\includegraphics[width=0.55\textwidth]{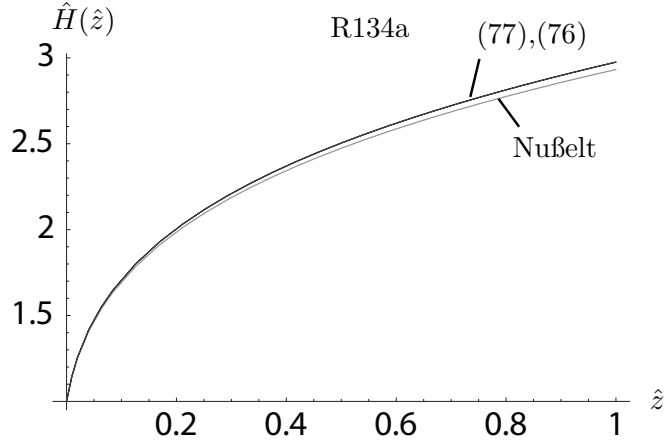}
\caption{ R134a: Nu{\ss}elt \eqref{odeNu}, without surface tension \eqref{ode}, with surface tension \eqref{odeST}\label{r134aST}}
\end{figure}

For R134a we assume the same tube geometry, initial film thickness, and temperature interval as before. The numerical solutions of~\eqref{ode}
and~\eqref{odeST} for R134a are shown in figure~\ref{r134aST}.
The R134a film is thinner than the water film. 
Beside this the results are similar as before. The difference between Nu{\ss}elt's solution and the numerical solutions of~\eqref{ode} and~\eqref{odeST} is smaller, because of the thinner film thickness of the R134a film.

\paragraph{Surface tension}

We could not observe an effect of surface tension on the film thickness for a vertical position of the tube.
Surface tension force acts normal to the interface. The effect of surface tension can be described by the tendency to minimize curvature (potential energy), and the minimal surface of the film is cylindrical.  Surface tension is the result of molecular forces. In the condensate flow the molecules are surrounded by other molecules and are attracted equally in all directions.

In a vertical tube the film cross sections are always circular, except at the tube inlet where the film thickness varies substantial over the tube length. The effect of surface tension to minimize curvature results in a more evenly distributed film thickness along the tube length.
An advantage of our model is that the tube radius does not need to be constant and an analysis of the model equations for condensation in a rotational tube where the radius is a function of tube length is straight forward.

\paragraph{Nu{\ss}elt number, dimensionless heat transfer}

\begin{figure}[h!]
\psfrag{Wasser u. R}{\hspace{-1cm}Water and R134a}\psfrag{134}{}\psfrag{a}{}
\psfrag{Nu}{$Nu$} \psfrag{z}{$\hat{z}$} \psfrag{wa}{Water}\psfrag{r1}{R134a}\centering
\leavevmode
\includegraphics[width=0.55\textwidth]{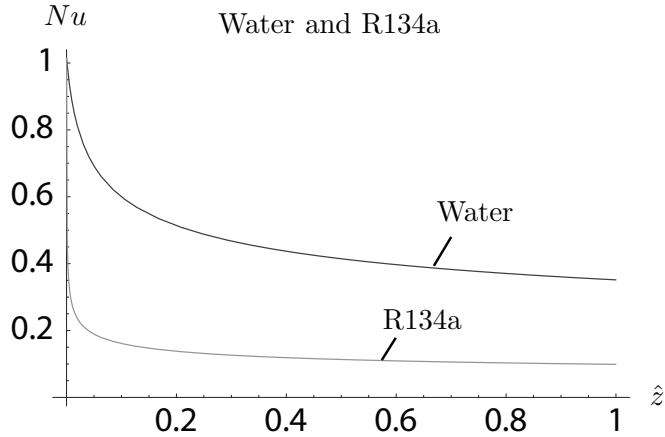}
\caption{ Nu{\ss}elt number of Water and R134a\label{alpha} }
\end{figure}

The heat transfer in the condensate film is mostly conductive so that the temperature profile
in the film is almost linear and we can write \beq q^{\prime\prime} = \alpha\,\Delta T\;. \eeq
Equating~$q^{\prime\prime}$ with the energy jump condition at the interface (where the heat transfer process occurs) gives
\beq
 \alpha\,\Delta T = \lambda\,\bs{\nabla}T\cdot\bs{n}=\dot{m}\,\Delta h_v\;,
\eeq
and in terms of dimensionless variables,
\beq
 \alpha\,\Delta T =
\frac{\lambda\,\Delta T}{H}\,\frac{\p\hat{T}}{\p\hat{r}} = \frac{\Delta
h_v\,\rho\,U\,H}{L}\,\hat{u}\,\frac{\p\hat{h}}{\p\hat{z}}\;. \eeq Multiplying the last
equation with~$\frac{H}{\lambda\,\Delta T}$ gives the local Nu{\ss}elt number, which is
defined as the dimensionless temperature gradient at the interface \beq \text{Nu} =
\frac{\alpha\,H}{\lambda} = \frac{\p\hat{T}}{\p\hat{r}} = \underbrace{\frac{\Delta
h_v\,\rho\,U\,H^2}{\lambda\,\Delta T\,L}}_{ \frac{\varepsilon\,\text{Pe}}{\text{ST}}}\,
\hat{u}\,\frac{\p\hat{h}}{\p\hat{z}}\;. \eeq Figure~\ref{alpha} shows the local Nu{\ss}elt number of Water and R134a as a function of tube length.
Note that the local Nu{\ss}elt number is almost inversely proportional to the film thickness.\\

Finally we compare our results with the mean Nu{\ss}elt number~$\overline{\text{Nu}}_{\text{Nu}}=(3\,\text{Re})^{1/3}$ for condensation along a flat plate according to Nu{ss}elt's theory and with the mean Nu{\ss}elt number given by Chen for co-current condensation in tubes~\cite{Bejan},
\begin{multline*}
\overline{\text{Nu}}_{\text{Chen}} = \left(
\text{Re}^{-0.44}\!\!+\!5.82\,10^{-6}\text{Re}^{0.8}\text{Pr}^{1/3} \right)^{1/2}\\
            +\left( 3.27\,10^{-4}\frac{\text{Pr}^{1.3}}{2\,R^2}
\left(\frac{\nu_w^2}{g}\right)^{2/3}\left(\frac{\eta_{wg}}{\eta_w}\right)^{0.156}\left(\frac{\rho_w^2}{\rho_w}\right)^{0.788}
 \frac{\text{Re}^{1.8}}{4.121}\right)^{1/2}\,.
\end{multline*}
Chen reviewed available experimental information for co-current condensation  inside
vertical tubes.
\begin{table}[h!]
\centering
\begin{tabular}{|l|c|c|}
\hline               & Water & R134a \\
\hline  $\overline{\text{Nu}}_{\text{Nu}}$  &   0.44 & 0.15  \\
 \hline $\overline{\text{Nu}}$      &   0.52 & 0.23  \\
\hline $\overline{\text{Nu}}_{\text{Chen}}$              &   0.74 & 0.37 \\
\hline
\end{tabular}\\[1ex]
\caption{Comparison of mean Nu{\ss}elt numbers\label{NusseltNumbers}}
\end{table}
His model is more accurate for Reynolds numbers greater than $\text{Re}=30$, since it takes
into account that at higher Reynolds numbers the film is thinner due to the co-current vapor
flow. This explains the difference between our model and Chen's model. Our model is based
on the physics of the problem and has a more general range of applicability.

\paragraph{Some remarks on inclined tubes}

We argued in section \ref{odesection} that the main forces in the condensation process are gravity force and surface tension force and that the position of the interface is mainly determined by the solution of the flow problem.

In an inclined tube gravity force can be divided into two components. One component acts in the plane of rotation and the other acts in the direction of the tube axis. Gravity causes the condensate to flow in a gathered stream at the bottom of the tube. Surface tension acts normally to the surface. Most likely for both fluids the balance of the two forces acting in the plane of rotation result in a cross section without rotational symmetry.

At the surface of the condensate the attraction forces result in a force that is directed inwards the condensate film such that the condensate film is pulled into circular shape. Surface tension force is the dominant force for water, but is comparably small for R134a, so that for a given inclination angle of the tube the cross section of the water condensate film will be more cylindrical and the R134a condensate film will mainly flow at the tube bottom.

The heat transfer through the condensate is mainly conductive. According to Fourier's law the conductive heat flux through a fluid for a given film thickness and a constant temperature difference is better if the film is thinner. So a better heat transfer can be expected for R134a, which would explain experimental results, \cite{Fiedler}. The model equations for condensation in an inclined tube will be partial differential equations.

\newpage
\begin{appendix}
\section{Material properties and dimensionless numbers\label{material}}
{\small
\textbf{Water:}\\[1ex]
Measured quantities (vapor has saturation temperature):\\[1ex]
\begin{tabular}{clrcllll}
\qquad& vapor temperature
& $T_s$ &=& 318.98 && K &\\
& temperature difference
& $T_s-T_w$ &$\approx$& 5 && K &\\
& pressure
& $ p $ &=& 10 & $10^3$& N\,m$^{-2}$ & \\
& mass flux
& $\dot{M}_l$ &=& 0.05 & $10^{-3}$ & kg\,s$^{-1}$ &  \\ 
&& $\dot{M}_v$ &=& 0.67 &$10^{-3}$ & kg\,s$^{-1}$ & \\ 
& initial film thickness
& $H_0$ & $\approx$ & 0.1 & $10^{-3}$ & m & \\
\end{tabular}\\[2ex]
Inner tube diameter, characteristic length, characteristic velocity, epsilon:\\[1ex]
\begin{tabular}{clcl}
\qquad
& $d_i$ = 2($R$ - $H_0$)& = & 7 - 0.2\,10$^{-3}$m \\
& $H = \tfrac{\D A_l}{\D\pi\, d_i}$ & = & 0.1\,$10^{-3}$  m  \\
& $U = \tfrac{\D\dot{M}_l}{\D\rho\, A_l}$ & = & 0.023\, m\,s$^{-1}$  \\
& $\varepsilon$  = $\tfrac{H}{L}$ & = &$\tfrac{0.1\,10^{-3}\text{m}}{0.5\,\text{m}} = 0.2\,10^{-3}$
\end{tabular}\\[2ex]
Material properties of water at $T_s$ and $p$:\\[1ex]
\begin{tabular}{clccllll}
& density
& $ \rho_l $ &=& 989.9 && kg\,m$^{-3}$ \\
&& $ \rho_v $ &=& 0.068 && kg\,m$^{-3}$ \\
& dynamical viscosity
& $ \mu_l $ &=& 0.60 & $10^{-3}$ & N\,s\,m$^{-2}$  \\
& $\mu=\nu\;\rho$ & $ \mu_v $ &=& 162.9  & $10^{-3}$ & N\,s\,m$^{-2}$ \\
& kinematical viscosity
& $ \nu_l $ &=& 0.606 & $10^{-6}$ & m$^2$\,s$^{-1}$ \\
&& $ \nu_v $ &=& 0.011 & $10^{-6}$ & m$^2$\,s$^{-1}$ \\
& thermal conductivity
& $ \lambda_l $ &=& 0.637 &  & W\,K$^{-1}$\,m$^{-1}$\\
&& $ \lambda_v $ &=& 19.98 & $10^{-3}$ & W\,K$^{-1}$\,m$^{-1}$ \\
& specific heat capacity
& $ c_l $ &=& 4.179 &$10^3$ & J\,kg$^{-1}$\,K$^{-1}$ \\
& latent heat of evaporation
& $ \Delta h_v $ &=& 2.393 & $10^3$ & J\,kg$^{-1}$ \\
& surface tension
& $ \sigma $ &=& 68.78 & $10^{-3}$ & N\,m$^{-1}$ \\
\end{tabular}\\[2ex]
Dimensionless numbers:\\[1ex]
\begin{tabular}{cllcllll}
\qquad & Re =$\tfrac{\D\dot{M}_l}{\D \mu\,\pi\,d_i}$~=~$\tfrac{\D\rho\,U\,H}{\D\mu} $ &=& 3.90 &\\
& Fr = $\tfrac{\D U^2}{\D g\, H}$ &=& 0.57 & \\
& Pr = $\tfrac{\D\mu\,c_P}{\D\lambda}$ &=& 3.94 \\
& Pe = Re\,Pr &=& 15.4 \\
& St = $\tfrac{\D c_P\,T_s}{\D\Delta h_v}$ &=& 557 &\\
& ST = $\tfrac{\D c_P\,\Delta T}{\D\Delta h_v}$ &=&  8.37 &\\
& We = $\tfrac{\D\rho\,H\,U^2}{\D \sigma}$ &=& 0.79\,$10^{-3}$\\
& Nu = $\tfrac{\D\alpha\,H}{\D\lambda}$ &&
\end{tabular}\\[2ex]

\textbf{R134a (1,1,1,2-Tetrafluorethan):}\\[1ex]
Measured quantities (vapor has saturation temperature):\\[1ex]
\begin{tabular}{clrcllll}
\qquad& vapor temperature
& $T_s$ &=& 297.15 && K &  \\
& wall temperature
& $T_w$ &=& 294.15 && K &  \\
\end{tabular}\\
\begin{tabular}{clrcllll}
& pressure
& $p$ &=& 0.65 & 10$^6$& kg\,m$^{-2}$ &\\
& condensate mass flux
& $\dot{M}_l$ &=& 0.367 & 10$^{-3}$ & kg\,s$^{-1}$ & \\
& film thickness (vertical tube)
& $H_0$ &=& 0.1 .. 0.2 & 10$^{-3}$& m & \\
\end{tabular}\\[2ex]
Characteristic length and characteristic velocity:\\[1ex]
\begin{tabular}{clcl}
\qquad
& $H = \frac{\D A_l}{\pi\, d_i}$ &=& 0.1\,10$^{-3}$ m \\
& $U = \frac{\D\dot{M}_l}{\D\rho\, A_l}$ &=& 0.142\, m\,s$^{-1}$ \\
\end{tabular}\\[2ex]
Material properties at $T_s$ and $p$ according to Tillner-Roth:\\[1ex]
\begin{tabular}{clcccllr}
& density
& $ \rho_l $ &=& 1210 && kg\,m$^{-3}$ \\
&& $ \rho_v $ &=& 31.39 && kg\,m$^{-3}$ \\
& dynamical viscosity
& $ \mu_l $ &=& 198.70 & 10$^{-6}$ & kg\,s\,m$^{-2}$  \\
&& $ \mu_v $ &=& 12.10  & 10$^{-6}$ & kg\,s\,m$^{-2}$ \\
& kinematical viscosity
& $ \nu_l $  &=& 0.164 & 10$^{-6}$ & m$^2$\,s$^{-1}$ \\     
&& $ \nu_v $  &=& 0.385 & 10$^{-6}$ & m$^2$\,s$^{-1}$ \\    
& thermal conductivity
& $ \lambda_l $ &=& 82.98 & 10$^{-3}$ & W\,K$^{-1}$\,m$^{-1}$ \\
&& $ \lambda_v $ &=& 14.35 & 10$^{-3}$ & W\,K$^{-1}$\,m$^{-1}$\\
& specific heat capacity
& $ c_l $ &=& 1.421 &  10$^3$ & J\,kg$^{-1}$\,K$^{-1}$ \\
&& $ c_v $ &=& 1.025 & 10$^3$ & J\,kg$^{-1}$\,K$^{-1}$ \\
& thermal diffusity
& $ a_l $ &=& 48.26 & 10$^{-9}$ & m$^2$\,s$^{-1}$           \\
&& $ a_v $ &=& 445.9 & 10$^{-9}$ & m$^2$\,s$^{-1}$\\
& latent heat of evaporation
& $ h_l $ &=& 233.1 &10$^3$ & J\,kg$^{-1}$ \\
&& $ h_v $ &=& 411.8 & 10$^3$ & J\,kg$^{-1}$ \\
&& $ \Delta h_v $ &=& 178.72 &10$^3$& J\,kg$^{-1}$ \\
& surface tension
& $ \sigma $ &=& 8.21 &  10$^{-3}$ & N\,m$^{-1}$
\end{tabular}\\[2ex]
Dimensionless numbers:\\[1ex]
\begin{tabular}{cllcll}
\qquad & Re = $\tfrac{\D\dot{M}_l}{\D \mu\,\pi\, d_i}$~=~$\tfrac{\D\rho\,U\,H}{\D\mu}$ &=& 86.46 &\\
& Fr = $\tfrac{\D U^2}{\D g\, D}$ &=& 20.55 & \\
& Pr = $\tfrac{\D\mu\,c_p}{\D\lambda}$ &=& 3.40 \\
& Pe = Re\,Pr &=& 294.2 \\
& St = $\tfrac{\D c_P\,T_s}{\D\Delta h_v}$ &=& 2.36 &\\
& ST = $\tfrac{\D c_P\,\Delta T}{\D\Delta h_v}$ &=& 0.040\\
& We = $\tfrac{\D\rho\,D\,U^2}{\D 2\,\sigma}$ &=& 0.297
\end{tabular}
}\\[3ex]
\end{appendix}

\end{document}